# On the accuracy of the Timoshenko beam theory above the critical frequency: best shear coefficient

J. A. Franco-Villafañe[*] and R. A. Méndez-Sánchez[†]

Instituto de Ciencias Físicas, Universidad Nacional Autónoma de México, A.P. 48-3, 62251 Cuernavaca, Morelos, México.

We obtain values for the shear coefficient both below and above the critical frequency by comparing the results of the Timoshenko beam theory with experimental data published recently. The best results are obtained, by a least-square fitting, when different values of the shear coefficient are used below and above the critical frequency.

1. Introduction

In recent years there has been renewed interest in the flexural vibrations of beams [1-4]. Even in the simplest case of a uniform beam, their vibrations are of great interest since not only is the validity of the theory still under discussion, but also the physical parameters that should be used within it. This is the case of the (Timoshenko) shear coefficient $\kappa$. This adjustment parameter appears in the Timoshenko beam theory, TBT, to estimate the shear force at the cross section of a beam in terms of the shear strain at the centroidal axis.

Until now there have been several theoretical studies attempting to obtain the best value for the shear coefficient [5-11]. Some commonly used values are $\kappa = 5/6$,

[*] Tel. +52 55 562 27788; Fax: +52 55 562 27775, Email addresses: jofravil@fis.unam.mx

[†] Tel. +52 55 562 27788; Fax: +52 55 562 27775, Email addresses: mendez@fis.unam.mx

$\kappa = \pi^2/12$, etc., for a beam with rectangular cross-section but there is no consensus about its value. Experimental studies, on the other hand, are scarce [12-14]. This is due to the fact that measurements with frequencies above the Bernoulli-Euler regime are needed. Numerical simulations, using finite elements analysis, have also been performed to calculate the shear coefficient [15].

To calculate the Timoshenko's shear coefficient it is assumed that the cross-sectional area is planar [16]. This hypothesis is correct at low frequencies but it seems to fail at frequencies higher than the critical frequency $f_c$. After a long debate, it was shown that in this regime, *i.e.* for $f > f_c$, $f$ being the frequency, the TBT is still valid [17] and that two families of normal modes appear as doublets [17-25]. One family of these doublets is called the ``second TBT spectrum''. In reference [17] it was also shown that the slope $\Psi$ along the cross-sectional area is no longer constant for the second TBT spectrum. Thus, it is expected that the value of the shear coefficient $\kappa$ changes for frequencies higher than $f_c$. A qualitative picture of the classification of the first and second TBT spectra, in terms of the shape of the cross-sectional area, is shown in Fig. 1.

Following the line of Ref. [14], in this paper we will give a new benchmark for the Timoshenko shear coefficient that is valid not only below but also above the critical frequency. In the next section the Timoshenko's beam theory with the different values of the shear coefficient is introduced. In Section 3 comparison is made of the different predictions of the TBT for different values of the shear coefficient $\kappa$ with recently published experiment results [17]. The article ends with some brief conclusions.

## 2. Timoshenko beam theory

The vertical displacement $\xi$ in the two-coefficient Timoshenko beam theory satisfies [7]

$$\frac{EI}{\rho A}\frac{\partial^4 \xi}{\partial z^4} - \frac{I}{A}\left(1+\frac{E}{\kappa_3 G}\right)\frac{\partial^4 \xi}{\partial z^2 \partial t^2} + \frac{\partial^2 \xi}{\partial t^2} + \frac{\rho I}{\kappa_1 GA}\frac{\partial^4 \xi}{\partial t^4} = 0, \qquad (1)$$

where $G$ and $E$ are the shear and Young's moduli, and $A$, $\rho$, and $I$ are the cross-sectional area, the density and the second moment of area, respectively. This one-dimensional theory predicts correctly the doublets [17] but assumes flat deformations of the cross-sectional area; any other deformation is absorbed into the shear coefficient $\kappa$.

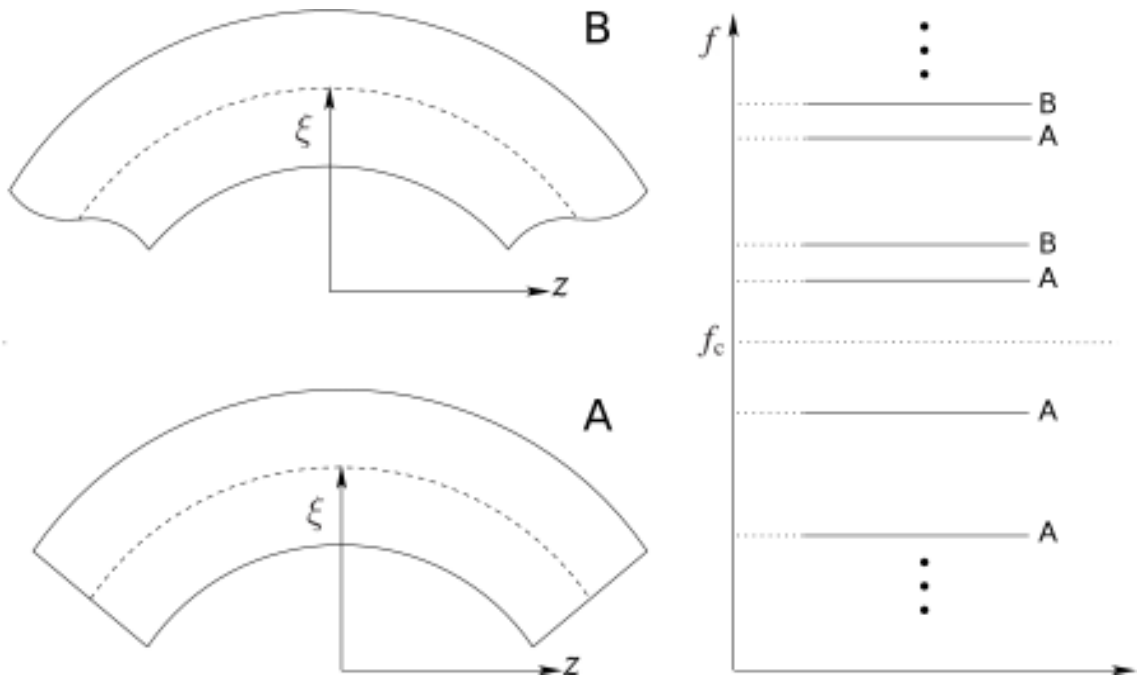

**Figure 1.** Displacement of the cross-sectional area for modes with frequencies below the critical frequency $f_c$ (A), for modes belonging to the first TBT spectrum (A), and for the second TBT spectrum (B).

When the beam is vibrating in a normal-mode, the separation of variables method can be used,

$$\xi(z,t) = e^{i\omega t} \chi(z),\tag{2}$$

where $\chi(z)$ is the time-independent displacement amplitude. With the previous *ansatz* the Timoshenko equation can be written as

$$\frac{d^4\chi}{dz^4} + \frac{\rho\omega^2}{M_r}\frac{d^2\chi}{dz^2} + \frac{\rho^2\omega^2}{\kappa_1 GE}\left(\omega^2 - \omega_c^2\right)\chi = 0,\tag{3}$$

where $M_r$ is the reduced modulus defined as

$$\frac{1}{M_r} = \frac{1}{E} + \frac{1}{\kappa_3 G},\tag{4}$$

and

$$\omega_c = 2\pi f_c = \sqrt{\frac{\kappa_1 GA}{\rho I}};\tag{5}$$

$f_c$ is known as the critical frequency.

For a free-free beam of length $L$ the boundary conditions, given by the vanishing of moments and shear forces, can be written in terms of the time-independent displacement amplitude only as [7,17,20,22]

$$\left.\frac{d^2\chi}{dz^2}\right|_{z=0,L} = -\left.\frac{\rho\omega^2 \chi}{\kappa_3 G}\right|_{z=0,L},\tag{6}$$

$$\left.\frac{d^3\chi}{dz^3}\right|_{z=0,L} = -\frac{\rho\omega^2}{M_r}\left.\frac{d\chi}{dz}\right|_{z=0,L}.\tag{7}$$

Notice that Eqs. (6) and (7) depend on the frequency and that they reduce to the Bernoulli-Euler boundary conditions, used in Ref. [14], when the terms on the right are much smaller than the respective terms at the left, *i.e.* for low frequencies.

The solution to Eq. (3), obtained using standard methods for ordinary differential equations, is given by

$$\chi(z) = a\exp(k_+ z) + b\exp(-k_+ z) + c\exp(k_- z) + d\exp(-k_- z), \quad (8)$$

where $a$, $b$, $c$, and $d$ are constants determined by the boundary conditions and

$$k_\pm^2(\omega) = \frac{\rho\omega^2}{2M_r}\left[-1 \pm \sqrt{1 - \frac{4M_r^2}{\kappa_1 GE}\left(1 - \frac{\omega_c^2}{\omega^2}\right)}\right]. \quad (9)$$

From the previous equations, it can be seen that the critical frequency separates the behavior of the solutions into two regimes: when $\omega < \omega_c$ there are two travelling waves and two exponential terms in Eq. (8) since the $k_-$ is imaginary and $k_+$ is real; when $\omega > \omega_c$ all the terms in Eq. (8) represent travelling waves; $k_\pm$ are both imaginary. A schematic behavior of $k_\pm$ is shown on Fig. 2. The frequency spectrum can be obtained by inserting Eq. (8) into the boundary conditions (6) and (7); the normal-mode frequencies are then obtained when $\det P = 0$ with

$$P = \begin{pmatrix} (k_+^2 + \frac{\rho\omega^2}{\kappa_3 G}) & (k_+^2 + \frac{\rho\omega^2}{\kappa_3 G}) & (k_-^2 + \frac{\rho\omega^2}{\kappa_3 G}) & (k_-^2 + \frac{\rho\omega^2}{\kappa_3 G}) \\ (k_+^2 + \frac{\rho\omega^2}{\kappa_3 G})e^{k_+ L} & (k_+^2 + \frac{\rho\omega^2}{\kappa_3 G})e^{-k_+ L} & (k_-^2 + \frac{\rho\omega^2}{\kappa_3 G})e^{k_- L} & (k_-^2 + \frac{\rho\omega^2}{\kappa_3 G})e^{-k_- L} \\ (k_+^3 + \frac{\rho\omega^2 k_+}{M_r}) & -(k_+^3 + \frac{\rho\omega^2 k_+}{M_r}) & (k_-^3 + \frac{\rho\omega^2 k_-}{M_r}) & -(k_-^3 + \frac{\rho\omega^2 k_-}{M_r}) \\ (k_+^3 + \frac{\rho\omega^2 k_+}{M_r})e^{k_+ L} & -(k_+^3 + \frac{\rho\omega^2 k_+}{M_r})e^{-k_+ L} & (k_-^3 + \frac{\rho\omega^2 k_-}{M_r})e^{k_- L} & -(k_-^3 + \frac{\rho\omega^2 k_-}{M_r})e^{-k_- L} \end{pmatrix}. \quad (10)$$

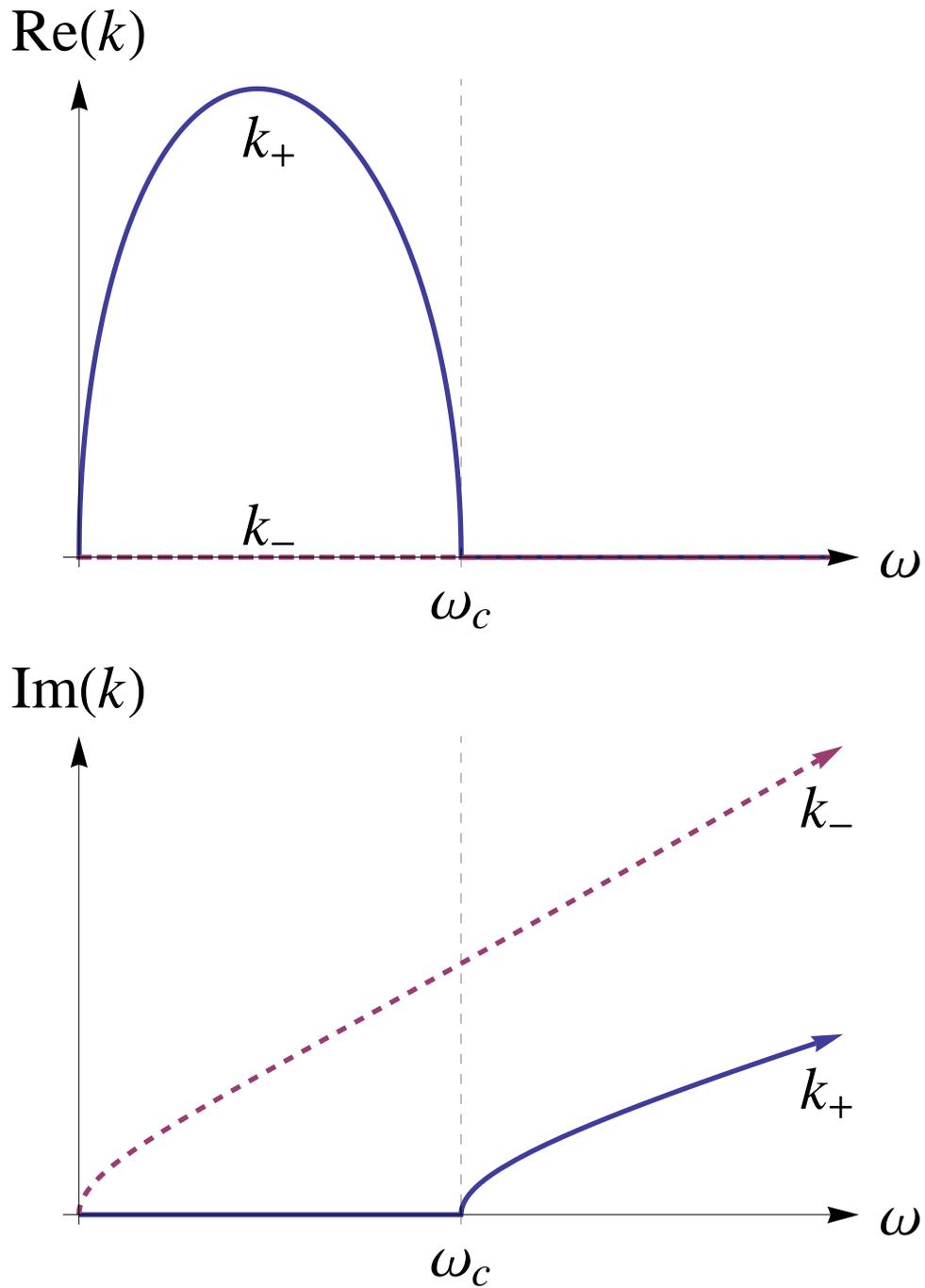

**Figure 2.** Schematic behavior of the wave vectors ($k_\pm$) as a function of the frequency $\omega$ for the Timoshenko's wave equation. The upper and lower panels correspond to the real and imaginary part of the wave vectors, respectively.

## 3. Results

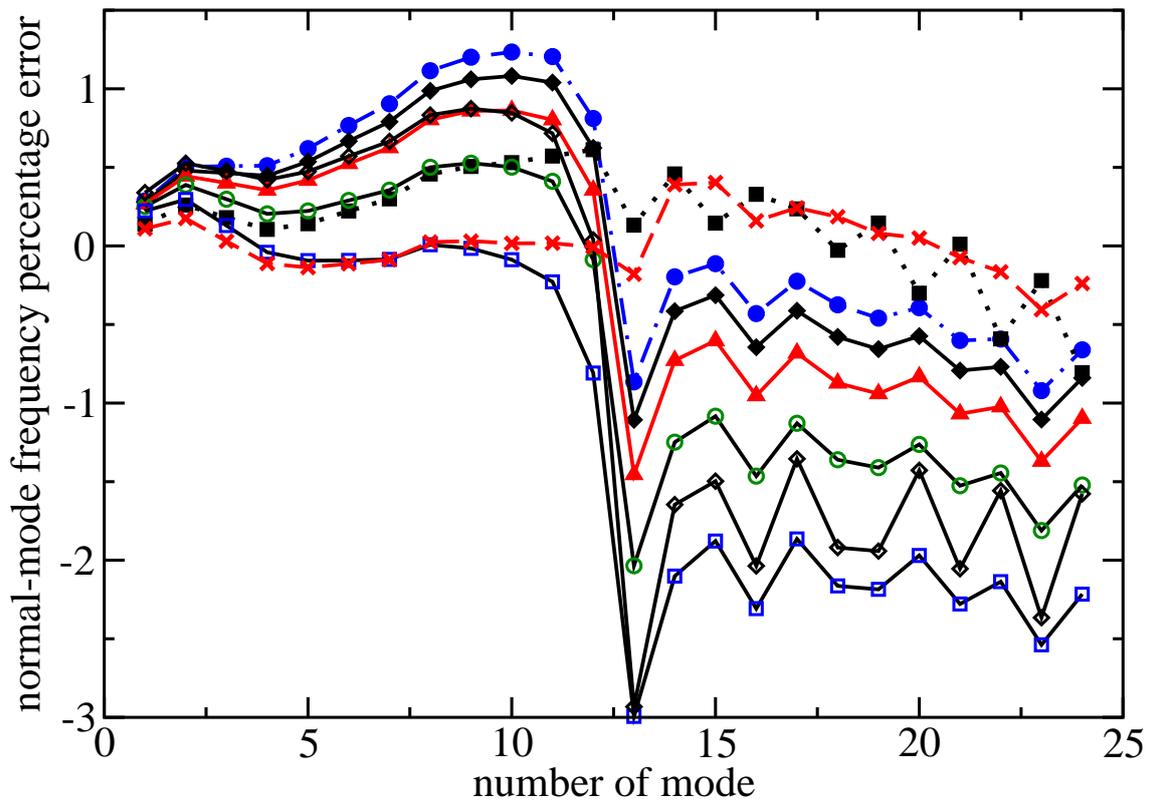

**Figure 3.** Percentage error in frequency between the experimental results of Ref. [17] and the theoretical predictions of the Timoshenko beam theory. The aluminum rod has a length $L = 0.500$ m, height $a = 0.0252$ m and width $b = 0.0504$ m; the elastic constants are $G = 26.92$ GPa, $E = 67.42$ GPa and $\rho = 2699.04$ kg/m$^3$. The plot markers correspond to those given in Table 1.

The lower 24 normal-mode frequencies of a beam of rectangular cross-section with length $L = 0.500$ m, height $a = 0.0252$ m and width $b = 0.0504$ m, were calculated by finding the roots of the determinant of Eq. (10). The elastic constants of the beam, taken from Ref. [17], are $G = 26.92$ GPa, $E = 67.42$ GPa and $\rho = 2699.04$ kg/m$^3$. Several values, found in the literature, for the Timoshenko shear coefficient ($\kappa = \kappa_1 = \kappa_3$) were used (cases I, II, III, IV, and V); these values are given in Table 1.

The results were compared with the experimental results of Ref. [17]. The error between these experimental results and the theory using two different values of $\kappa_1$ and $\kappa_3$ is plotted in Fig. 3. As can be seen in this figure, excellent agreement between the theory and the experiment is found at low frequencies; one can also notice that the error grows with the frequency. Close to $f_c$, which varies between 196187 Hz and 202383 Hz for the different cases reported here (see Eq. (5)), the error between the different theories and the experiment grows up to 3 %; the larger errors are found close to $f_c$, around mode 13, where a peak in the error is observed. In this figure one can also observe that the results above the critical frequency present larger errors than the results below $f_c$.

| Case | Symbol | Timoshenko shear coefficient(s) |
| --- | --- | --- |
| I | ◆ | $\kappa = 5/6$ |
| II | ▲ | $\kappa = 0.83945$ |
| III | ○ | $\kappa = 10(1+\nu)/(12+11\nu)$ |
| IV | □ | $\kappa = 5(1+\nu)/(6+5\nu)$ |
| V | ◇ | $\kappa_1 = 10(1+\nu)/(12+11\nu)$, $\kappa_3 = 5(1+\nu)/(6+5\nu)$ |
| VI | ● | $\kappa = 0.8291$ |
| VII | ■ | $\kappa_1 = 0.80811$, $\kappa_3 = 0.83292$ |
| VIII | × | below $f_c$: $\kappa_1 = 0.82003$, $\kappa_3 = 0.84651$ <br> above $f_c$: $\tilde{\kappa}_1 = 0.81687$, $\tilde{\kappa}_3 = 0.81923$ |

**Table 1.** Different values of the Timoshenko coefficients used to test the TBT. The values used in cases VI, VII and VIII were calculated using least squares (see text).

We also calculated the best shear coefficients using least squares in three different ways: with one independent coefficient $\kappa = \kappa_1 = \kappa_3$ (case VI); with $\kappa_1$ and $\kappa_3$ as independent coefficients (case VII); and, due to the change of regime introduced by the critical frequency, with four coefficients, $\kappa_1$ and $\kappa_3$ for $f < f_c$ and $\widetilde{\kappa}_1$ and $\widetilde{\kappa}_3$ for $f > f_c$ (case VIII). As can be seen from Fig. 3, case VIII, the error is smaller than 0.5 % for all resonances below and above the critical frequency. The error is smaller than 0.62 % in case VII.

### 4. Conclusions.

In this paper it has been shown that the predictions of Timoshenko's beam theory can be very accurate not only below but also above the critical frequency $f_c$. This was acheived by comparing the theoretical results for different values of the Timoshenko shear coefficient with experimental results recently published. When the two–coefficient Timoshenko beam theory is used, away from the critical frequency, the difference between theory and experiment is smaller than 0.5%. This difference is smaller than 0.62% if one includes the results that are close to the critical frequency. The results strongly suggest that the value of the shear coefficient above the critical frequency is different from the value below it.

### 5. Acknowledgments.

This work was supported by DGAPA-UNAM under project IN103115. RAMS is indebted to J. Flores, G. Monsivais, A. Morales, L. Gutierrez, and A. Díaz-de-Anda, for useful discussions. JAFV acknowledge the support of CONACyT under project CB-2010/154586. We acknowledge the kind hospitality of Centro Internacional de Ciencias A. C. for group meetings celebrated frequently there.